\documentclass[twocolumn]{autart}

\usepackage{graphics} % for pdf, bitmapped graphics files
\usepackage{epsfig} % for postscript graphics files
\graphicspath{ {images/} }
\usepackage{color,soul} % for highlighting
\usepackage{enumitem}  

\usepackage{times} % assumes new font selection scheme installed
\usepackage{cite}

\usepackage{amsmath} % assumes amsmath package installed
\usepackage{amssymb}
\usepackage{url}
\renewcommand{\rm}[1]{{\color{red} \st{#1}}}
\begin{document}
% 
% \title{\LARGE \bf 
% Discord in Bayesian Quickest Change Detection: When will numeric issues emerge?}

% \author{Jason J. Ford and others% <-this % stops a space
% \thanks{The authors are with the School of Electrical Engineering and Computer Science, Queensland University of Technology, 2 George St, Brisbane QLD, 4000 Australia. 
% {\tt\small  j2.ford@qut.edu.au}.}%
% % {\tt\small jasmin.james@connect.qut.edu.au, j2.ford@qut.edu.au, t.molloy@qut.edu.au}. 
% % This work was supported by funding from the Australian Research Council Centre of Excellence CE140100016 in Robotic Vision.}%
% }

% % make the title area
% \maketitle

\begin{frontmatter}

\title{On the Informativeness of Measurements in Shiryaev's Bayesian Quickest Change Detection} % Title, preferably not more 

%\thanks[footnoteinfo]{This paper was not presented at any IFAC 
%meeting. Corresponding author M.~T.~Cicero. Tel. +XXXIX-VI-mmmxxi. 
%Fax +XXXIX-VI-mmmxxv.}

\author[qut]{Jason J. Ford}\ead{j2.ford@qut.edu.au},    % Add the 
\author[qut]{Jasmin  James}\ead{jasmin.james@connect.qut.edu.au},               % e-mail address 
\author[qut]{Timothy L. Molloy}\ead{t.molloy@qut.edu.au}  % (ead) as shown

\address[qut]{School of Electrical Engineering and Computer Science, Queensland University of Technology, 2 George St, Brisbane QLD, 4000 Australia. }  % Please supply                                              
% \address[Rome]{Senate House, Rome}             % full addresses
% \address[Baiae]{The White House, Baiae}        % here.

                             % keyword list or with the 

\begin{abstract}
%In this paper we identify conditions under which 
This paper provides the first description of a weak practical super-martingale phenomenon that can emerge in 
%the complement of 
the test statistic
%monotonic posteriors
%, in the form of monotonic posteriors, 
%emerges %(before the change event) 
in Shiryaev's Bayesian quickest change detection (QCD) problem.  
%After establish bounds on the posteriors, 
We establish %the emergence of 
that this super-martingale phenomenon %as 
can emerge under a condition on the relative entropy between pre and post change densities when the measurements are insufficiently informative to overcome the change time's geometric prior.
%and the no change posterior can become a weak super-martingale.
%s as a condition on
%that the 
%characterizes the  via a condition on pre and post change models lead to 
%under . 
% Under this condition, Shiryaev's problem becomes ill-posed in the sense that measurements lose informativeness.
%are insufficiently informative.
%the test statistic is dominated by the change-time's prior.
%, as the no-change posterior becomes a practical exponential-supermartingale and both the posteriors become monotonic.
We illustrate this super-martingale phenomenon in a simple Bayesian QCD problem 
%with pre and post change unit variance Gaussian densities having different means 
which highlights the unsuitability of
%this  in unavoidable and 
%that Bayesian QCD is potentially 
Shiryaev's test statistic for detecting subtle change events.
\end{abstract}
\end{frontmatter}
%%%%%%%%%%%%%%%%%%%%%%%%%%%%%%%%%%%%%%%%%%%%%%%%%%%%%%%%%%%%%%%%%%%%%%%%%%%%%%%%%%%%%%%%%
\section{Introduction}
Quickly detecting a change in the statistics of a process is an important signal processing problem with application in a diverse range of areas including: automatic control
\cite{basseville1993}, quality control\cite{Hwang,Lai1998,basseville1993},
%signal processing \cite{Vaswani}, 
statistics \cite{TartakovskyPFA}, target detection\cite{Ru,Lai2013} and many more \cite[Ch, 1.3]{Tartakovsky2014}. % for an overview of change point detection's rich history. %fault detection \cite{Hwang} and automatic control \cite{basseville1993}. 
%Statistical change detection problem can be posed for both non-Bayesian \hl{ ref for LORDON etc} and Bayesian change events. 
In the classic Bayesian quickest change detection (QCD) problem, it is assumed that a permanent change in the statistics of an observed process occurs at some random time 
(see \cite[Ch. 1.2]{Tartakovsky2014} for a comparison with non-Bayesian QCD).
%(see \cite{krishnamurthy2016,James2018} for recent Bayesian QCD generalizations and see \cite[Ch. 1.2]{Tartakovsky2014} for a comparison with non-Bayesian QCD). 
%he classic Bayesian QCD problem has since been generalised (cf.  \cite{krishnamurthy2016,James2018}) and its original formulation and solution provide a template for the solution of these new problems.
% In the Bayesian quickest change detection (QCD) problem, it is assumed that a permanent change in the statistics of an observed sequence of random variables occurs at some random time, see \cite[Ch, 1.2]{Tartakovsky2014} for a history of Bayesian and non-Bayesian QCD techniques and \cite{krishnamurthy2016,James2018} for some recent extensions of Bayesian QCD problems. 
%The change time is assumed to have a geometric prior, and 
The classic Bayesian QCD objective is to minimize the average detection delay subject to a constraint on the probability of a false alarm. When the change time has a geometric prior, Shiryaev established the optimal stopping rule as a test of whether the change posterior probability is above a threshold \cite{Shiryaev}.
This paper investigates the properties of Shiryaev's famous test statistic in weak measurement environments.
% , and 
% %In this paper 
% we identify conditions under which the Shiryaev Bayesian quickest change detection problem is ill-posed due to the measurement models being insufficient informative.

%\hl{Maybe a list of things known, leading to what is not known last sentence to transition to this papers focus. - Would be easiest for TM to write.}

% In practical applications of Bayesian QCD tec
% There is less known about practical aspects of QCD 

The main contribution here is to provide the first report and characterization of a super-martingale phenomenon in Shiryaev's Bayesian QCD problem (see \cite{Lai1998} and \cite{Polunchenko} for extensive investigations of martingale phenomenon in 
%the test statistics of 
other QCD rules).
%have been extensively investigated 
%, classic Bayesian QCD problem has not been previously studied. %this paper is the first reporting and characteristing a phenomenon in the classic Bayesian QCD problem.
This paper introduces a new weak practical super-martingale concept
% and exploits the recent maximal inequality for non-negative supermartingales established in \cite{TM2017}
and exploits the maximal inequality for non-negative supermartingales
to characterise of conditions under which the Bayesian QCD %problem is ill-posed in the sense that 
measurements are not sufficiently informative and Shiryaev's test statistic is dominated by the change time's prior. Interestingly, the identified super-martingale phenomenon appears suddenly once  an information theoretic requirement on the pre and post change densities holds (rather than emerging as a graceful degradation). Practically, in applications with weak measurements, these observations motivate consideration of %alternative techniques
%the use of Potential remedies in applications with weak measurements involve 
subtle problem adjustments, such as in the quickest intermittent signal detection problem \cite{James2018} which generalizes Shiryaev's problem for use in a vision-based aircraft detection application, or using non-Bayesian QCD such as the Lorden criterion \cite{Lorden}.

%  Recently, alternatives to Shiryaev Bayesian QCD problem have proposed that work well in 

% Potential remedies in applications with weak measurements involve subtle model adjustments, such as the  quickest intermittent signal detection problem \cite{James2018} which generalizes Shiryaev's problem, or using non-Bayesian QCD such as the Lorden criterion \cite{Lorden}. 
%Finally, we would expect similar  phenomenon to arise in filter problems involving non-ergodic hidden Markov models with weak observations.

The specific contributions are:
\begin{enumerate}[label=(\roman*)]
\item Establishing a condition in terms of the change time's geometric prior and the relative entropy between  {pre and post change} densities that identifies when measurements are insufficiently informative.
\item Establishing that when measurements are insufficiently informative,
 Shiryaev's test statistic can exhibit a super-martingale phenomenon; that is, the log of no change posterior is a weak practical super-martingale.
%exhibit monotonic discord (specifically, the \hl{no-change} posterior %becomes a practical exponential-supermartingale). % and %not informative in s well posed QCD problem. 
\item Providing an example exhibiting this super-martingale phenomenon to illustrate a situation where Shiryaev's  Bayesian QCD approach is potentially unsuitable for detecting subtle change events.
\end{enumerate}
We would expect similar phenomenon to emerge in recent Bayesian QCD generalizations involving non-ergodic models \cite{krishnamurthy2016}.

% There is very little established about this sort of discord, but physics community have recently identified a discorded in HMM conditional mean estimates.  The context here is different, and arises because the signal model under the Bayesian QCD problem is not-ergodic.

% Possible literature review:
% HMM discord papers in physics community, Applications problems (DAA i.e. ergodicity issues (wrapping and then ISD as solution), Speech absorbing left to might models
%To achieve these contributions we introduce a new weak practical super-martingale concept.
%To achieve these results a new maximal inequality tech

%%%%%%%%%%%%%%%%%%%%%%%%%%%%%%%%%%%%%%%%%%%%%%%%%%%%%%%%%%%%%%%%%%%%%%%%%%%%%%%%%%%%%%%%%
\section{Shiryaev's Bayesian Quickest Change Detection Problem and Optimal Solution}\label{sec:probForm}
For $k >0$, let $y_k \in \mathcal{Y}$ be an %{\it i.i.d.} 
independent and individually distributed ({\it i.i.d.})
sequence of random variables taking values in the set $\mathcal{Y} \subseteq \mathbb{R}^M$. Initially, the random variables $y_k$ have a pre change (marginal) probability density $b^1 (\cdot)$ before, at some random change time $\nu \ge 1$, switching to having a post change (marginal) probability density $b^2 (\cdot)$. We will assume for $i \in \{1,2\}$, that $b^i(.)< B$ for some finite $B<\infty$.
%for $\mu_{i-1}<k<\nu_i$ and  $f_k^{2}(\cdot)$ for $\nu_i <k < \mu_i$. 
%In this paper we consider an i.i.d. random process $y_k \in E_Y$ whose mean switches at random time instants. Specifically for $k \geq 0$, let $y_k$ be a sequence of i.i.d. random variables whose density switches between $f^1(y) = \psi(y-d_1)$ and $f^2(y) = \psi(y-d_2)$ where $\psi(\cdot)$ is a density and $d_1$ and $d_2$ are two constants which represent a change in mean. For example $\psi(\cdot)$ could be a zero mean Gaussian distribution with known variance $\sigma^2$. 
% \hl{This problem setup needs to be re-written as appropriate for Bayesian quickest change problem.}
For $k \ge 0$, let random variable $X_k \in  \{e_1, e_2\}$ denote 
%the change event through a sequence of random variables representing a 
a change event process in the sense that $X_k=e_1$ for $k < \nu$  
%denotes no-change has occurred 
and $X_k=e_2$ for $k \ge \nu$. % denotes the change has occurred. 
Here $e_i \in R^2$ are indicator vectors with 1 in the $i$th element, and zero elsewhere.  
Let $y_{[1,k]}\triangleq \{y_1, \ldots, y_k \} $ be shorthand for measurement sequences.

% and $a$ is the probability of self transition for  $e_2$.  

% We consider measurement sequence $y_k \in Y \subseteq R$ where described by the conditional density functions $b^i(y_k) \triangleq P(y_{k}  | X_k = e_i)$. We assume, for $i=1,2$, $b^i(,)< B$ for some $B<\infty$.

% We consider the situation of finite alphabet observations $Y_k \in E_Y = \{f_1, f_2\}$, whose output probabilities at each time instant are given by $B^{i,j}_k \triangleq P(Y_{k} = e_i | X_k = e_j)$ for $i,j \in \{1,2\}$ as
% \begin{equation}
% B = \begin{bmatrix}
% b_1 & 1- b_2 \\
% 1-b_1 & b_2
% \end{bmatrix}.
% \end{equation}
% where we assume that $b_1>0.5$ and $b_2>0.5$ so that the measurements are meaningfully associated with the corresponding state.

Before we formally state Shiryaev's Bayesian QCD problem, 
%Quickest Change Detection problem, 
 let us first introduce a probability measure space. 
Let $\mathcal{F}_k=\sigma(y_{[1,k]})$ denote the filtration generated by $y_{[1,k]}$. We will assume the
existence of a probability space $(\Omega,\mathcal{F},P_\nu)$ where
$\Omega$ is a sample space of sequences $y_{[1,\infty]}$, $\sigma$-algebra $\mathcal{F}=\cup_{k=1}^\infty \mathcal{F}_k$ with the convention that $\mathcal{F}_0=\{0,\Omega\}$, and
$P_\nu$ is the probability measure constructed using Kolmogorov's extension on the joint probability density function of the observations
$
p_\nu(y_{[1,k]})=\Pi_{i=1}^{\nu-1}b^1 (y_i) \Pi_{j=\nu}^{k}b^2 (y_j)
$ where we define $\Pi_{j=\nu}^{k}b^2 (y_j)=1$ when $k<\nu$. We will let $E_\nu$ denote expectation under $P_\nu$ and use the probability measure $P_\infty$ and expectation $E_\infty$ to denote the special case when there is no change event. We let $D\left(b^1(y_k)\big|\big|{b^2(y_k)}\right)  \triangleq E_\infty \left[\log\left( \frac{b^1(y_k)}{b^2(y_k)} \right)    \right]$ denote the relative entropy between pre and post change densities.

In Bayesian QCD problem the change time $\nu \ge 1$ that $X_k$ transitions from $e_1$ to $e_2$ is considered to be an unknown random variable with  prior distribution $\pi_k\triangleq P(\nu=k)$ for $k\ge 1$. This allows us to construct a new averaged measure $P_\pi(G)=\sum_{k=1}^{\infty} \pi_k P_k(G)$ for all $G \in \mathcal{F}$ and we let $E_\pi$ denote the corresponding expectation operation.  In Shiryaev's problem we consider the special case of the geometric prior $\pi_k=(1-\rho)^{k-1} \rho$ for some $\rho \in ( 0,1)$ (and set $\pi_k\triangleq 0$, $k<1$). 

Let $\tau > 0$ be a stopping time with respect to filtration $\mathcal{F}_k$.
We can now introduce the Shiryaev cost criterion \cite{Shiryaev} to trade-off average detection delay with probability of false alarm %(PFA) 
as
\begin{equation}
J(\tau) = \ c E_\pi \bigg[(\tau-\nu)^+\bigg] +  P_\pi  (\tau< \nu), 
\end{equation}
where $(\tau-\nu)^+ \triangleq \max(0,\tau -\nu)$, $c$ is the delay penalty and the problem is to minimise $\inf_{\tau} J(\tau)$.  

For $i \in \{1,2\}$, let the  no change  and change  posterior probabilities be denoted $\hat{X}_k^i \triangleq P_{\pi}(X_k=e_i | y_{[1,k]})$, respectively.
%noting that we can write $\hat{X}_k^2=1-\hat{X}_k^1$. 
Noting that we can write $\hat{X}_k^2=1-\hat{X}_k^1$, allows us to write
Shiryaev's optimal stopping rule for this cost criterion 
%can be written 
in terms of the no change posterior probability as
\[
\tau=\inf\left\{k \ge 1 :  \hat{X}^1_k < 1- h  \right\},
%\tau=\inf\left\{k \ge 1 :  \hat{X}^2_k > h  \right\},
\]
where $h$ is a threshold selected to control the probability of false alarm, as it can be shown that the probability of false alarm satisfies
$ P_\pi (\tau < \nu)< 1-h$ 
%PFA$(\tau) \le 1-h$ 
\cite{James2018}.  
%Alternatively, Shiryaev's optimal stopping rule can be rewritten in terms of the no change 
% \[
% \tau=\inf\left\{k \ge 1 :  \hat{X}^1_k < 1- h  \right\},
% %\tau=\inf\left\{k \ge 1 :  \hat{X}^2_k > h  \right\},
% \]

\section{The Emergence of the Super-Martingale Phenomenon}
To develop conditions under which the test statistic $\hat{X}^1_k$ of Shiryaev's rule exhibits rapid decrease even in the no change regime, we first
introduce the following result that establishes how to efficiently calculate it. 
%implement Shiryaev's rule through calculation of the no-change posterior probability  $\hat{X}^1_k$.
%%LEMMA
\begin{lem} \label{lem:filter}
For $k>0$, given a sequence of measurements $y_{[1,k]}$ 
%$\{y_1,\ldots,y_k\}$
the no change posterior probability $\hat{X}^1_k$
%$\hat{X}_k^1 \triangleq P(X_k=e_1 | y_{[1,k]})$ 
is given by the scalar recursion
\begin{equation}
\hat{X}_k^1=N_k (1-\rho) b^1(y_k)  \hat{X}_{k-1}^1     
\label{equ:hmmfa}
\end{equation}
with $\hat{X}_0^1=1$ and  the {normalization factor} 
\begin{equation}
N_k^{-1}   =    b^2(y_k) + 
%\nonumber \\
%&&
(1-\rho) \left( b^1(y_k)  -  b^2(y_k)  \right)\hat{X}_{k-1}^1.
\label{equ:hmmNa}
\end{equation}
\end{lem}

%%PROOF
\begin{pf}
As defined above, $X_k$ is a first order time-homogeneous Markov chain whose  transition probabilities at each time instant are given by $A^{i,j} \triangleq P(X_{k+1} = e_i | X_k = e_j)$ for $i,j \in \{1,2\}$ as
\begin{equation}
A = \begin{bmatrix}
1-\rho & 0 \\
\rho & 1
\end{bmatrix},
\end{equation}
where   $X_0=e_1$, and $X_k$ is observed via the random variables $y_k$. %are a measurement process for the hidden state $X_k$. 
Hence, the no change posterior $\hat{X}_k^1$ can efficiently be calculated by hidden Markov model filter \cite{elliott1995}, where $\hat{X}_k=[\hat{X}_{k}^1,\hat{X}_{k}^2]'$,
\[
\hat{X}_k= N_k \mbox{diag}( [b^1(y_k) ,  b^2(y_k)] ) A \hat{X}_{k-1}
\]
where $N_k= \langle \mbox{diag}( [b^1(y_k) ,  b^2(y_k)] ) A \hat{X}_{k-1} , \underline{1} \rangle ^{-1}$ and with $\hat{X}_0=e_1$.
Noting that $\hat{X}_k^2=1-\hat{X}_k^1$,
%(with $\hat{X}_k^2 \triangleq P(X_k=e_2 | y_{[1,k]})$), 
then simple algebra lets us write  \eqref{equ:hmmfa}. Then we note that
\begin{eqnarray*}
N_k^{-1}&=& b^1(y_k)  \hspace*{-0.1em} \left( \hspace*{-0.1em} (1  \hspace*{-0.1em}-  \hspace*{-0.1em}\rho) \hat{X}_{k-1}^1  \hspace*{-0.1em} \right)  
%\nonumber \\
%&&
\hspace*{-0.3em} + \hspace*{-0.2em} b^2(y_k) \hspace*{-0.2em} \left( \hspace*{-0.1em} \rho \hat{X}_{k-1}^1 \hspace*{-0.2em}  + \hspace*{-0.1em} (1
 \hspace*{-0.1em}-  \hspace*{-0.1em}\hat{X}_{k-1}^1)  \hspace*{-0.1em} \right) 
\nonumber \\
& = & b^1(y_k)  \left((1-\rho) \hat{X}_{k-1}^1 \right)  
%\nonumber \\
%&&
+ b^2(y_k) \left( (\rho-1) \hat{X}_{k-1}^1 + 1  \right) 
\nonumber \\
& = &  b^2(y_k) + 
%\nonumber \\
%&&
(1-\rho) (b^1(y_k) - b^2(y_k))\hat{X}_{k-1}^1
\end{eqnarray*}
giving \eqref{equ:hmmNa}. 
%Setting $a=1$, gives \eqref{equ:hmmfa} and \eqref{equ:hmmNa}.
This completes the proof.
\end{pf}
 
 To facilitate characterization of our test statistic's behaviour let us introduce %$M_k  \triangleq N_k (1-\rho) (BY_k)^1 \in R$ 
$M_k  \triangleq N_k (1-\rho) b^1(y_k) \in R$, noting that we can write $\log(\hat{X}_{k}^1)=\log(M_k) + \log(\hat{X}_{k-1}^1)$,
%, and 
%note that $\log(\hat{X}_{k}^1)=\log(M_k) + \log(\hat{X}_{k-1}^1)$,
% % so the update factor at time $k$ in \eqref{equ:hmmfa}. 
% Recursion \eqref{equ:hmmfa} can now be written as 
% %$\hat{X}_k^1 = M_k \hat{X}_{k-1}^1$; noting $\hat{X}_{0}^1=1$ leads to the no-change posterior descriptions 
% $\hat{X}_k^1 = \Pi_{\ell=1}^k M_\ell$ or equivalently $\log(\hat{X}_k^1) = \Sigma_{\ell=1}^k \log(M_\ell)$. 
% % This notation provides a compact way to investigate the properties of the no-change posterior (and hence indirectly properties of Shiryaev's test statistic $\hat{X}_{n }^2$). 
%We can now 
and establish the following %$\hat{X}_k$ dependent exponential 
bound on $\log(M_k)$.  
%then idempotent property 

\begin{lem} \label{lem:bound}
%Consider $a=1$ and assume $b_1=b_2=b$. 
($\hat{X}_k$ dependent bound on $\log(M_k)$)
For  any $\delta > 0$, there is a $h_\delta>0$ such that for any $\hat{X}_{k-1}^1< h_\delta$ we have 
\[ 
%M_L \le 
E_\pi \left[\log(M_k) \Big| \hat{X}_{k-1}^1  \right] < \log(1-\rho) + 
%(2b -1) \log\left(\frac{b }{1-b }\right)   
D\left(b^1(y_k)\big|\big|{b^2(y_k)}\right) 
%D_\pi ( b^1(y_k)||b^2(y_k) )
+ \delta.
\]  
\end{lem}

% %%PROOF
\begin{pf}  
% We first 
% note, 
% for any $i,j \in \{ 1,2\}$, that
% $E_\pi\left[ \log \left(  b^i(y_k) \right) \langle {X_k}, e_j \rangle \Big|    \hat{X}_{k-1}^1  \right] = E_\pi \left[ \log\left(  b^i(y_k) \right)  \langle {X_k}, e_j \rangle \right]$
% %\]
% because $b^i(y_k)$ is independent of $\hat{X}_{k-1}^1$ when $X_k$ is known. Hence, because $\langle {X_k}, e_1 \rangle+\langle {X_k}, e_2 \rangle =1$, we have
% \begin{eqnarray}
% E_\pi\left[ \log\left(  b^i(y_k) \right)   \Big|    \hat{X}_{k-1}^1  \right]  
% &=& E_\pi\left[\log\left(  b^i(y_k) \right)  \left(\langle {X_k}, e_1 \rangle+\langle {X_k}, e_2 \rangle \right)  \Big|    \hat{X}_{k-1}^1   \right] 
% \nonumber \\ 
% &=& E_\pi\left[ \log\left(  b^i(y_k) \right) \right].
% \label{equ:nocon}
% \end{eqnarray}
% \[
% E^1\left[ \log\left(  b^i(y_k) \right)   \Big|    \hat{X}_{k-1}^1  \right]  =E^1\left[ \log\left(  b^i(y_k) \right) \right].
% \] 
We define  
\begin{eqnarray*}
&&\gamma_k  \triangleq   \log  \left( b^2(y_k) \right) \\
&&  \hspace*{3em} - \log   \left(   b^2(y_k)   + (1 - \rho)(b^1(y_k)  -   b^2(y_k) )\hat{X}_{k-1}^1   \right).
\end{eqnarray*}
% $\gamma_k \hspace*{-0.2em} \triangleq \hspace*{-0.2em} \log \hspace*{-0.2em} \left( b^2(y_k) \right) - \log \hspace*{-0.2em} \left( \hspace*{-0.2em} b^2(y_k)  \hspace*{-0.2em} + \hspace*{-0.2em} (1\hspace*{-0.2em} -\hspace*{-0.2em} \rho)(b^1(y_k) \hspace*{-0.2em} - \hspace*{-0.2em} b^2(y_k) )\hat{X}_{k-1}^1 \hspace*{-0.2em} \right)$.
%and then note that we can write $\log (N_k)  =- \log ( b^2(y_k)) + \gamma_k$.
Using \eqref{equ:hmmNa} lets us write
\begin{eqnarray}
 E_\pi \left[  \log (N_k) \Big|    \hat{X}_{k-1}^1  \right] &=& 
\nonumber \\
&& \hspace*{-9em} - E_\pi\left[   \log \left( b^2(y_k) \hspace*{-0.2em} + \hspace*{-0.2em} (1\hspace*{-0.2em} -\hspace*{-0.2em} \rho)(b^1(y_k) \hspace*{-0.2em} -\hspace*{-0.2em} b^2(y_k) )\hat{X}_{k-1}^1 \right)\Big|    \hat{X}_{k-1}^1  \right] 
 \nonumber \\
 && \hspace*{-9em}    = 
- E_\pi \left[ \log\left(  b^2(y_k) \right) \Big|    \hat{X}_{k-1}^1 \right] +  E_\pi \left[ \gamma_k  \Big|  \hat{X}_{k-1}^1  \right].
\label{equ:nk}
\end{eqnarray}
It then follows from  %\eqref{equ:nocon}, 
\eqref{equ:nk} and the definition of  $M_k$ that
\begin{eqnarray}
E_\pi[\log(M_k) | \hat{X}_{k-1}^1 ]  &=& 
 \log(1-\rho)  
\nonumber \\
&& \hspace*{-9em} 
+ E_\pi\left[ \log(N_k)\Big|  \hat{X}_{k-1}^1 \right] +  E_\pi\left[ \log( b^1(y_k) ) \Big| \hat{X}_{k-1}^1  \right]
\nonumber \\
% &=& 
% \log(1-\rho)  + 
% E_\pi\left[  \log (N_k) \Big|    \hat{X}_{k-1}^1  \right]  +  E_\pi\left[ \log( b^1(y_k) )  \right] 
%  \nonumber \\
&& \hspace*{-9em}  =  \log(1-\rho)  + 
E_\pi\left[ \log( b^1(y_k) ) \Big| \hat{X}_{k-1}^1 \right] 
\nonumber \\
&& \hspace*{-9em} 
- E_\pi \left[ \log\left(  b^2(y_k) \right) \Big| \hat{X}_{k-1}^1 \right] + E_\pi \left[ \gamma_k  \Big|  \hat{X}_{k-1}^1  \right] \nonumber \\
&& \hspace*{-9em}  =  \log(1    -  \rho)   +  
E_\pi  \left[ \log\left(   \frac{b^1(y_k)}{ b^2(y_k)}   \right)   \Big| \hat{X}_{k-1}^1    \right] 
%- E_\pi \left[ \log\left(  b^2(y_k) \right) \Big| \hat{X}_{k-1}^1 \right] 
%D_\pi ( b^1(y_k)||b^2(y_k) )
\hspace*{-0.3em}  
\nonumber \\
&& \hspace*{-5em} 
+ 
  E_\pi   \left[ \gamma_k  \Big|  \hat{X}_{k-1}^1   \right].
% &\le &  \log(1-\rho)  + 
% E \left[ \log\left(  b^2(y_k) \right) \right] +  E\left[ \log( b^1(y_k) )  \right] + \gamma.
\label{equ:logm}
\end{eqnarray}
% where the 2nd line follows from \eqref{equ:nocon}, and the 3rd following from
% and simple algebra given the lemma result.
Noting that $\log(x)$ is a continuous (monotonic increasing) in $x$  and that $b^i(.)<B$ are finite gives that for any $\delta > 0$, there is a $h_\delta>0$  such that for any $\hat{X}_{k-1}^1< h_\delta$ we have
$E_\pi \left[ \gamma_k  \Big| \hat{X}_{k-1}^1  \right] \le \delta$,  
%Hence 
and
therefore \eqref{equ:logm} gives that 
%for any $\delta > 0$, there is a $h_\delta>0$, such that for any $\hat{X}_{k-1}^1< h_\delta$ we have
\begin{eqnarray}\label{equ:cbound}
E_\pi[\log(M_k) | \hat{X}_{k-1}^1 ]  & < &    \log(1-\rho)  
\nonumber \\
&&  \hspace*{-1em} +  
E_\pi   \left[ \log\left(  \frac{b^1(y_k)}{ b^2(y_k)}  \right)    \Big| \hat{X}_{k-1}^1  \right]   +   \delta.
\end{eqnarray}
% \begin{equation}\label{equ:cbound}
% E_\pi[\log(M_k) | \hat{X}_{k-1}^1 ] \hspace*{-0.2em}  < \hspace*{-0.2em}   \log(1-\rho) \hspace*{-0.1em}  
% \nonumber \\
% &&  \hspace*{-0.2em} + \hspace*{-0.1em} 
% E_\pi\hspace*{-0.3em}  \left[ \log\left(\hspace*{-0.2em}  \frac{b^1(y_k)}{ b^2(y_k)} \hspace*{-0.2em} \right) \hspace*{-0.2em}  \Big| \hat{X}_{k-1}^1 \hspace*{-0.2em} \right] \hspace*{-0.1em}  +\hspace*{-0.05em}  \delta.
% \end{equation}
% %Simple algebra gives the lemma result.
% Let 
% %$D(p||q)\triangleq E[p\log(\frac{p}{q})]$,
% $D(p(y_k)||q(y_k))\triangleq \int_{\mathcal{Y}}p(y_k)\log(\frac{p(y_k)}{q(y_k)})d y_k$
% %denote the Kullback-Leiber distance, %then idempotent property 
% % %gives
Then using $P_\pi(y_k| \hat{X}_{k-1}^1 )= P_\pi(X_k=e_1| \hat{X}_{k-1}^1 ) P_\pi(y_k|X_k=e_1, \hat{X}_{k-1}^1) +P_\pi(X_k=e_2| \hat{X}_{k-1}^1 ) P_\pi(y_k|X_k=e_2, \hat{X}_{k-1}^1 )$ and
 $P_\pi(X_k=e_1|\hat{X}_{k-1}^1 )=(1-\rho)\hat{X}_{k-1}^1$ 
 gives 
\begin{eqnarray}\label{equ:ebound}
E_\pi\left[  \log\left( \frac{b^1(y_k)}{ b^2(y_k)} \right)  \Big| \hat{X}_{k-1}^1 \right] =
% &=&   E_\pi\left[ \langle {X_k}, e_1 \rangle 
% \frac{b^1(y_k)}{ b^2(y_k)}  \right] + E_\pi\left[ \langle {X_k}, e_2 \rangle  \frac{b^1(y_k)}{ b^2(y_k)}  \right]
\nonumber\\
&& \hspace*{-12em}   (1-\rho)\hat{X}_{k-1}^1   \int_{\mathcal{Y}} b^1(y_k)\log\left( \frac{b^1(y_k)}{ b^2(y_k)} \right) d y_k 
\nonumber\\
&& \hspace*{-12em} 
+ (1-(1-\rho)\hat{X}_{k-1}^1 )  \int_{\mathcal{Y}}  b^2(y_k)\log\left( \frac{b^1(y_k)}{ b^2(y_k)} \right) d y_k 
\nonumber\\
&&\hspace*{-12em}  =   (1-\rho)\hat{X}_{k-1}^1  E_\infty\left[\log\left( \frac{b^1(y_k)}{ b^2(y_k)} \right)  \right] 
\nonumber\\
&& \hspace*{-12em} - (1-(1-\rho)\hat{X}_{k-1}^1 )  \int_{\mathcal{Y}}  b^2(y_k)\log\left( \frac{b^2(y_k)}{ b^1(y_k)} \right) d y_k 
\nonumber\\
&&\hspace*{-12em}  =   (1-\rho)\hat{X}_{k-1}^1 D\left(b^1(y_k)\big|\big|{b^2(y_k)}\right) 
\nonumber\\
&& \hspace*{-12em}- (1-(1-\rho)\hat{X}_{k-1}^1 )  E_0\left[\log\left( \frac{b^2(y_k)}{ b^1(y_k)} \right)  \right] 
\nonumber\\
% &=&  P_\pi(X_k=e_1)  E_\pi\left[ b^1(y_k)\log\left( \frac{b^1(y_k)}{ b^2(y_k)} \right)  \right] + (P_\pi(X_k=e_1)-1) E_\pi\left[ b^2(y_k)\log\left( \frac{b^2(y_k)}{ b^1(y_k)} \right)   \right]
% \nonumber\\
% &=&  P_\pi(X_k=e_1)  \left[ D(b^1(y_k)||b^2(y_k)) +D(b^2(y_k)||b^1(y_k))\right]  - D(b^2(y_k)||b^1(y_k)).
&&\hspace*{-12em}<   D\left(b^1(y_k)\big|\big|{b^2(y_k)}\right) 
\end{eqnarray}
% \begin{eqnarray*}
% E_\pi\left[  \log\left( \frac{b^1(y_k)}{ b^2(y_k)} \right) \right]
% % &=&   E_\pi\left[ \langle {X_k}, e_1 \rangle 
% % \frac{b^1(y_k)}{ b^2(y_k)}  \right] + E_\pi\left[ \langle {X_k}, e_2 \rangle  \frac{b^1(y_k)}{ b^2(y_k)}  \right]
% % \nonumber\\
% &=&  P_\pi(X_k=e_1)  E_\pi\left[ b^1(y_k) \log\left( \frac{b^1(y_k)}{ b^2(y_k)} \right)  \right] + (1-P_\pi(X_k=e_1)) E_\pi\left[ b^2(y_k) \log\left( \frac{b^1(y_k)}{ b^2(y_k)} \right) \right]
% \nonumber\\
% &=&  P_\pi(X_k=e_1)  E_\pi\left[ b^1(y_k)\log\left( \frac{b^1(y_k)}{ b^2(y_k)} \right)  \right] + (P_\pi(X_k=e_1)-1) E_\pi\left[ b^2(y_k)\log\left( \frac{b^2(y_k)}{ b^1(y_k)} \right)   \right]
% \nonumber\\
% &=&  P_\pi(X_k=e_1)  \left[ D(b^1(y_k)||b^2(y_k)) +D(b^2(y_k)||b^1(y_k))\right]  - D(b^2(y_k)||b^1(y_k)).
% \end{eqnarray*}
because $ (1-\rho)\hat{X}_{k-1}^1 < 1$, $(1-(1-\rho)\hat{X}_{k-1}^1 )\ge 0$ and $E_0\left[\log\left( \frac{b^2(y_k)}{ b^1(y_k)} \right)  \right] \ge 0$.
% Similarly we can write
% \[
% E_\pi\left[  \log\left( \frac{b^1(y_k)}{ b^2(y_k)} \right) \Big| \hat{X}_{k-1}^1 \right] =  P_\pi(X_k=e_1 |\hat{X}_{k-1}^1 )  \left[ D(b^1(y_k)||b^2(y_k)) +D(b^2(y_k)||b^1(y_k))\right]  - D(b^2(y_k)||b^1(y_k)).
% \]
% Noting $P_\pi(X_k=e_1) = (1-\rho)^k$ and $P_\pi(X_k=e_1|\hat{X}_{k-1}^1 )=(1-\rho)\hat{X}_{k-1}^1$ then these two relationships show that if $\hat{X}_{k-1}^1 < (1-\rho)^{k-1}$ then
% \begin{equation}
% E_\pi\left[  \log\left( \frac{b^1(y_k)}{ b^2(y_k)} \right) \Big| \hat{X}_{k-1}^1 \right] < E_\pi\left[  \log\left( \frac{b^1(y_k)}{ b^2(y_k)} \right) \right].
% \label{equ:ebound}
% \end{equation}
% Finally we note that $E_\pi \left[\log(M_k) \Big| \hat{X}_{k-1}^1  \right]$ is time-invariant (depends only on the value of $\hat{X}_{k-1}^1$) and hence the $\hat{X}_{1}^1 < (1-\rho)$ establishing the
% bound \eqref{equ:ebound} holds at $k=2$, means the bound similarly holds for at any $k$ when $\hat{X}_{k-1}^1 < (1-\rho)$. Select $h_\delta$ to the smaller of $(1-\rho)$ and the value needed for \eqref{equ:cbound} to hold.
Substitution of \eqref{equ:ebound} into \eqref{equ:cbound} %and setting $h_\delta=(1-\rho)^{k-1}$
gives the lemma result.
\end{pf}
%where $P( X_k |\hat{X}_{k-1}^1 )=A \hat{X}_{k-1}^1$.

Recall that we can write $\log(\hat{X}_{k}^1)=\log(M_k) + \log(\hat{X}_{k-1}^1)$. Hence
Lemma \ref{lem:bound} provides a bound on the test statistic increment $\log(M_k)$ which
%shows that the increment in $\log(\hat{X}_{k}^1)$ can be bounded. 
%Whilst $E_\pi \left[ \log\left(  b^2(y_k) \right) \right]>0$ the fact that $\log(1-\rho)<0$ raises the possibility that $\log(\hat{X}_{k}^1)$ is decreasing.
%The Lemma \ref{lem:bound} bound result 
allows us to investigate conditions under which the measurements are insufficient to overcome the geometric prior information $\log(1-\rho)<0$, and $\log(\hat{X}_k^1)$ becomes a weak practical super-martingale in the following sense:
% which an interval trap emerges in which 
% %\hl{monotonic posteriors} 
% $\log(\hat{X}_k^1)$
% becomes a weak super-martingale prior to the change event.
% %to possibly emerge ; that is, 
% %if $\hat{X}_{k}^1$ is sufficient small, then 
% % it is monotonic decreasing and $\hat{X}_{k}^1$ is monotonic
% %increasing in the absence of measurement support.  
% For this purpose we introduce a %internal ball-trap 
% weak practical super-martingale concept in the following definition.
\begin{defn} (Weak practical super-martingale) If for any arbitrarily small 
$\delta_p>0$ 
there exists a 
%$h_s \in (0,h_\delta)$ 
$h_s>0$
such that if $ \hat{X}_{k}^1 < h_s $ then
% \[
% P_\pi \left(\mbox{for all } n\ge k, E_\pi  [\log(\hat{X}_{n+1}^1)|\log(\hat{X}_{n}^1) ]  < \log(\hat{X}_{n}^1) \right)
% > 1-\delta_p.
% \]
\begin{eqnarray*}
&& P_\pi \hspace*{-0.2em} \left( \hspace*{-0.1em} \mbox{for all } n\ge k, E_\pi  [\log(\hat{X}_{n+1}^1)| \hspace*{-0.1em}\log(\hat{X}_{n}^1) ] \hspace*{-0.2em} < \hspace*{-0.2em} \log(\hat{X}_{n}^1) \hspace*{-0.2em} \right)
\\
&& \hspace{19em} > 1-\delta_p.
\end{eqnarray*}
and the log of the no change posterior probability $\log(\hat{X}_{k}^1)$ 
%is practically monotonic decreasing in the sense of being 
is called a weak practical super-martingale.
\label{def:pssm}
\end{defn}
%Because $\log(x)$ is monotonic increasing in argument x, 

We now establish our   theorem which provides conditions under which measurements are insufficiently informative and this super-martingale 
phenomenon emerges.
%$\hat{X}_{k}^1$ is a
%practical exponential-supermartingale. 
%monotonic posteriors ``trap" exists.

\begin{thm} \label{thm:pssm}
(Insufficiently informative measurements)
If the relative entropy between probability densities ${b^1(\cdot)}$ and ${b^2(\cdot)}$ is sufficient small, namely
\begin{equation}
%M_L \le 
%E_\infty \left[\log\left( \frac{b^1(y_k)}{b^2(y_k)} \right)    \right] 
D\left(b^1(y_k)\big|\big|{b^2(y_k)}\right) 
%+ \delta 
< \log(1/(1-\rho))
%- \log(1-\rho) 
%(2b -1) \log\left(\frac{b }{1-b }\right)   
\label{equ:REbound}
\end{equation}
%holds for any $\delta>0$, 
then the measurements are insufficiently informative in  %there is a $h_\delta>0$ such that for $\hat{X}_{k}^1< h_\delta$
the sense that
$\log(\hat{X}_{k}^1)$ is a weak practical super-martingale (cf. Definition \ref{def:pssm}).
%  $\log(\hat{X}_{k}^1)$ is practically a (strict) exponential-supermartingale in the sense that for any arbitrarily small 
% $\delta_p>0$ 
% there exists a $h_c \in (0,h_\delta)$ such that if $ \hat{X}_{k}^1 < h_c $ for some $k$ then
% \[
% P\left(\mbox{for all } n\ge k, E [\log(\hat{X}_{n+1}^1)|\log(\hat{X}_{n}^1) ]  < \log(\hat{X}_{n}^1) \right) > 1-\delta_p.
%\]
\end{thm}
\begin{pf}
% First note that  $E_\infty \left[\log\left( \frac{b^1(y_k)}{b^2(y_k)} \right)    \right]  \ge E_\pi \left[\log\left( \frac{b^1(y_k)}{b^2(y_k)} \right) \right]$ for all $k>0$, from the definition $ E_\pi$ and the observation that 
% $E_\infty \left[\log\left( \frac{b^1(y_k)}{b^2(y_k)} \right)    \right]  \ge E_\nu \left[\log\left( \frac{b^1(y_k)}{b^2(y_k)} \right) \right]$ for all $k,\nu>0$,
% %\hl{Do we need to add assumption/clarify ? - this assume measurements are more like to come from the correct density.}
From Lemma \ref{lem:bound}, the bound \eqref{equ:REbound} gives that 
%Lemma \ref{lem:bound} holds
%combined with Lemma \ref{lem:bound} implies that if $b \in E(\rho,\epsilon)$ then 
%for any $\delta<\epsilon$, https://www.overleaf.com/17155202tctbpjddjxpp#
%and 
there exists a $h_\delta>0$ such that for $\hat{X}_{k-1}^1 < h_\delta$ we have $E_\pi[\log(M_k)| \hat{X}_{k-1}^1]  < 0$ and hence that
$\log(\hat{X}_{k}^1)$ satisfies the super-martingale property
\begin{equation}
E_\pi [\log( \hat{X}_{k}^1) |  \log( \hat{X}_{k-1}^1) ] < \log( \hat{X}_{k-1}^1)
\label{equ:down}
\end{equation}
noting $ E_\pi[\log( \hat{X}_{k}^1) |  \log( \hat{X}_{k-1}^1)]  = E_\pi [\log(M_k)| \log( \hat{X}_{k-1}^1)]   
%+ E[ \log(\hat{X}_{\ell-1}^1 ) | {X}_\ell^1=1] 
+ \log( \hat{X}_{k-1}^1)$ and that conditioning on $\log( \hat{X}_{k-1}^1)$ and $\hat{X}_{k-1}^1$ are equivalent.
% $E[\log( \hat{X}_{k}^1) |   \hat{X}_{k-1}^1] = E[\log( \hat{X}_{k}^1) |  \log( \hat{X}_{k-1}^1) ]$ allows us to write  
It remains to establish if $\hat{X}_{k-1}^1$ remains trapped in $[0,h_\delta)$ or escapes.

%To understand the condition \eqref{equ:REbound}, 
%To establish the practical strict exponential-supermartingale result 
Let us introduce %$h_p=\alpha h_\delta$ and 
$h_s \triangleq \beta h_\delta$ and $h_m \triangleq \beta \eta h_\delta$, with some $ \beta, \eta<1$ as bounding parameters to manage our possibly unbounded $\log( \hat{X}_{k-1}^1)$ super-martingale process. 
We define a new process
%two new processes $\bar{X}_{k}^1 \triangleq \min(\hat{X}_{k}^1,h_\delta)$ and  
%$Z_k \triangleq \max( \log(\bar{X}_{k}^1/h_m),0)$,
$Z_k \triangleq \max( \log( \hat{X}_{k}^1/h_m),0)$.
% $Z_k \triangleq \max(E[\log(\bar{X}_{k}^1)|\hat{X}_{k-1}^1]-\log(\delta_{MP}),0)$ 
%where $\delta_{MP}\in (0,h_\delta)$ is an arbitrarily small number. 
We now note that \eqref{equ:down} gives that $Z_k$ is a non-negative super-martingale and hence by the maximal inequality for non-negative super-martingales (cf. \cite[Lemma 1]{TM2017})
we have, for any $k$. 
that
%The best reference I have is prop 7 in http://galton.uchicago.edu/~lalley/Courses/385/Martingales.pdf
\[
P_\pi\left( \max_{n \ge k }  Z_n  \ge (\log(h_\delta/h_m))\right) \le \frac{E_\pi[Z_{k}]}{\log(h_\delta/h_m)}.
\]
Noting that $Z_n \ge  \log(\hat{X}_{n}^1/h_m)$ and that if $\hat{X}_{k}^1 < h_s$ then $E_\pi[Z_{k}]<\log(h_s/h_m)$ gives
\[
P_\pi\left( \max_{n \ge k }  \log( \hat{X}_{n}^1)  \ge \log(h_\delta) \right) %\le \frac{h_s}{\log(\hat{X}_{k}^1/\delta_{MP})} 
<\frac{\log(h_s/h_m )}{ \log(h_\delta/h_m)}. %< h_s/\log(2)
%B(h_s,\delta_{MP},\hat{X}_{k}^1)
%\frac{h_s}{\log(\hat{X}_{k}^1)-\log(\delta_{MP})}
\] 
%After noting $\max_{n \ge k } E[ \log(\hat{X}_{n}^1) ] > E[ \log(\hat{X}_{k+1}^1) ]$, 
Rewriting in terms of the complimentary set for the maximal event gives, if $\hat{X}_{k}^1 < h_s$ that
\[
P_\pi\left( \max_{n \ge k }  \log( \hat{X}_{n}^1)  < \log(h_\delta)  \right) %\le \frac{h_s}{\log(\hat{X}_{k}^1/\delta_{MP})} 
> 1- \delta_l
%\frac{\log(h_s/\delta_{MP}))}{ \log(h_p/\delta_{MP})} %< h_s/\log(2)
%B(h_s,\delta_{MP},\hat{X}_{k}^1)
%\frac{h_s}{\log(\hat{X}_{k}^1)-\log(\delta_{MP})}
\] 
where $\delta_l \triangleq\log(h_s/h_m))/\log(h_\delta/h_m)$ can be written as
$\delta_l=\log(\eta)/\log(\beta \eta)$.
% we note that
% \[
% \delta_l = \frac{\log(\gamma)}{\log(\beta \gamma)}.
% \] 
We note that the event $\max_{n \ge k }  \log(\hat{X}_{n}^1)  < \log(h_\delta)$ implies for all $n \ge k$ that $\hat{X}_{n}^1 < h_\delta$ and hence by Lemma \ref{lem:bound} that \eqref{equ:down}
%event $E[\log( \hat{X}_{n+1}^1) |  \log( \hat{X}_{n }^1) ] < \log( \hat{X}_{n }^1)$ 
holds for all $n \ge k$. 
The theorem result follows by noting that for any $\delta_p>0$ we can find a $\beta$ (or equivalently a $h_s>0$) so that $\delta_l \le \delta_p$ and the
%follows that for any choice of $\delta_p>0$ there is a $\gamma>0$  such that the Theorem statement using 
Definition \ref{def:pssm} property holds. % (noting $h_\delta>\delta_{MP}$ and both are fixed) so that the lemma result holds.
% After writing $h_c=h_s/\log(h_\delta/\delta_{MP})$, %and setting $k=\ell-1$ 
% it then follows, for any $\delta_p>0$, there is a $h_c>0$ such that the final lemma result holds.
\end{pf}

%, if reaching the interval $\hat{X}_{k}^1 < h_s$. 

Theorem \ref{thm:pssm} establishes that unless the relative entropy between pre and post change densities $D\left(b^1(y_k)\big|\big|{b^2(y_k)}\right)$ is sufficiently large, %then the measurements may not 
%able to overcome the change times geometric prior, and 
the no change posterior $\hat{X}_{k}^1$ is a weak practical super-martingale under 
%the no change measure 
$E_\pi$
%definition %$\hat{X}_{k}^1$ being a 
%provides a description of 
and hence there exists a trap defined by the interval $\hat{X}_{k }^1 < h_\delta$ 
% %then the interval $\hat{X}_{k}^1 < h_s$ becomes a trap 
% which triggers the described stochastic exponential tendency of $\hat{X}_{k}^1$ to decrease %and the related Shiryaev's test statistic $\hat{X}^2_k$ stochastic tendency to 
% become 
where Shiryaev's test statistic  becomes
increasingly confident that the change has occurred even if it has not.  
%Further, as noted in \cite{TartakovskyPFA},  the probability of Shiryaev's rule issuing a false alarm in finite time is 1; hence the interval trap is reached in finite time in the sense that $P_\pi(n< \infty: \hat{X}_{n}^1 < h_s)=1$.
Further, we note that on sufficiently long sequence of measurements there is non zero probability of entering the interval $\hat{X}_{k }^1 < h_\delta$. 
%(whenever 
%%$E_\infty \left[\log\left( \frac{b^1(y_k)}{b^2(y_k)} \right)    \right]$ 
%the relative entropy is finite, there is non-zero probability of sequence of measurements such that $\hat{X}_{k }^1$  %\eqref{equ:hmmfa} 
%decreases enough).
A test statistic that can exhibit such incorrect increasing confidence on non-pathological sequences is problematic in a practical setting and hence we interpret the existence of this interval trap under the condition of Theorem \ref{thm:pssm} as meaning %that %Shiryaev's problem is  {ill-posed} when
the measurements are insufficiently informative. 
To understand the behaviour of Shiryaev's rule and the role of the relative entropy $D\left(b^1(y_k)\big|\big|{b^2(y_k)}\right)$ % (and size of the interval trap) 
first consider the limit case $b^1(\cdot)=b^2(\cdot)$. In this case, $D\left(b^1(y_k)\big|\big|{b^2(y_k)}\right)$ is zero, the posterior is given by $\hat{X}_k^1=(1-\rho)^k$, Shiryaev's rule becomes the deterministic rule
%to trigger %once $(1-\rho)^k < (1-h)$; that is, 
%at
to stop at the
earliest time at or after
%or immediately after 
$\log(1-h)/\log(1-\rho)$
%$\tau=\log(1-h)/\log(1-\rho)$ 
and $h_\delta=1$. Informally, a similar geometric prior $(1-\rho)^k$ mechanism is driving the 
super-martingale phenomenon that occurs when $D\left(b^1(y_k)\big|\big|{b^2(y_k)}\right)$ is non-zero but less than $\log(1/(1-\rho))$, with $h_\delta \in (0,1)$.
%is non-zero but the measurements are insufficiently informative to overcome the 
Finally, we note that as $D\left(b^1(y_k)\big|\big|{b^2(y_k)}\right)$ increases towards the critical value of $\log(1/(1-\rho))$ then $h_\delta$ decreases towards $0$, and the probability of entering the trap interval decays.

\section{Example: Bayesian Quickest Change Detection With Gaussian Densities}\label{sec:sim}
\begin{prop} \label{lem:setgaussuan}
%Consider $a=1$ and assume $b_1=b_2=b$. 
Consider %an finite alphabet observation model $Y=\{e_1,e_2\}$ 
Shiryaev's Bayesian quickest change detection problem with
pre and post change (marginal) probabilities densities 
%Consider an unit variance Gaussian observation model $Y\in R$ model observation probabilities 
$b^1(y_k)=\frac{1}{2 \pi} \exp(-y_k^2/2)$ and $b^2(y_k)=\frac{1}{2 \pi} \exp(-(y_k-m)^2/2)$.
Consider the set
\begin{equation}
\mathcal{M}(\rho) \triangleq   \left\{ m: \frac{m^2}{2}  < \log(1/(1-\rho)) \right\}.
\label{equ:set}
\end{equation}
%for some $\delta>0$.
%If the set $E(\rho,\epsilon)$ is non-empty for a given $\epsilon>0$, 
% For any given $\rho>0$, 
% there exists a $\delta>0$ such that 
$\mathcal{M}(\rho)$ is non-empty. Further, there exists a $m_c$ such that $\mathcal{M}(\rho)$
has a threshold structure in the sense of $\mathcal{M}(\rho)=\left\{m : 0 < m< m_c\right\}$, where 
%$m_c \ge \sqrt{2(\log(1/(1-\rho))-\delta)}$.
$m_c = \sqrt{2\log(1/(1-\rho))}$.
%that may depend on  $\delta$ and $\rho$.  
Finally, when $m \in \mathcal{M}(\rho)$,
then the measurements are insufficiently informative in  %there is a $h_\delta>0$ such that for $\hat{X}_{k}^1< h_\delta$
that
$\log(\hat{X}_{k}^1)$ is a weak practical super-martingale (cf. Definition \ref{def:pssm}).
% the measurements are insufficiently informative in the 
% sense $\log(\hat{X}_{k}^1)$ is practically a (strict) exponential-supermartingale in sense for any arbitrarily small 
% $\delta_p>0$ 
% there exists a $h_c \in (0,h_\delta)$ such that if $ \hat{X}_{k}^1 < h_c $ for some $k$ then
% \[
% P\left(\mbox{for all } n\ge k, E [\log(\hat{X}_{n+1}^1)|\log(\hat{X}_{n}^1) ]  < \log(\hat{X}_{n}^1) \right) > 1-\delta_p.
% \]
\end{prop}

\begin{pf}
To establish %there exists a $\delta>0$ such 
that $\mathcal{M}(\rho)$ is non-empty we note that for any $\rho>0$  there exist a $\epsilon>0$ 
such that $\log \left(1/(1-\rho)\right) > \epsilon$.
As $\lim_{m \rightarrow 0} \left[\frac{m^2}{2} \right]=0$
then for any $\epsilon>0$, there must be at least one $m>0$   such that $\frac{m^2}{2}  < \epsilon$ and hence %there exists a $\delta>0$ having 
this $m>0$ as an element of the non-empty $\mathcal{M}(\rho)$.
% To show elements are described by an interval we note that, for any $0 < \beta \le 1$
% \begin{eqnarray}
% \frac{\beta m^2}{2} & \le & 
% \frac{ m^2}{2}
% \nonumber  \\
% \label{equ:beta}
% \end{eqnarray}
The interval result  follows by noting that if $m \in \mathcal{M}(\rho)$  then $\beta m \in \mathcal{M}(\rho)$ for all $0 < \beta \le 1$, and this means the set $\mathcal{M}(\rho)$ can be described as the interval $\left\{m : 0 < m< m_c\right\}$ with some critical largest element $m_c$. Algebra and the monotonic increasing nature of $m^2/2$ gives that $m_c=\sqrt{2\log(1/(1-\rho))}$. 
The final result follows from Lemma \ref{lem:bound} and noting that the relative entropy between these two Gaussians is given by $\frac{m^2}{2}$ \cite[Example 4.1.9]{basseville1993} .
% \begin{eqnarray}
% E \left[\log\left( \frac{b^1(y_k)}{b^2(y_k)} \right)    \right] &=& \int \log\left( \frac{b^1(y)}{b^2(y)} \right) b^1(y) dy\\
% &=& \int \log\left(  \frac{\frac{1}{2\pi}\exp(-y^2/2)}{\frac{1}{2\pi}\exp(-(y-m)^2/2)} \right) b^1(y) dy\\
% &=& \int \left(-y^2/2+(y-m)^2/2 \right)  b^1(y) dy\\
% &=& \frac{1}{2} \int m^2b^1(y) dy - \frac{1}{2}\int2my b^1(y)  dy \\
% &=& \frac{m^2}{2}
% \nonumber
% \end{eqnarray}
% where in the last line we have used $\int b^1(y)dy=1$ and $\int y b^1(y) = E[b^1(y)]=0$. 
\end{pf}

\begin{figure}
\begin{center}
\includegraphics[scale=0.4]{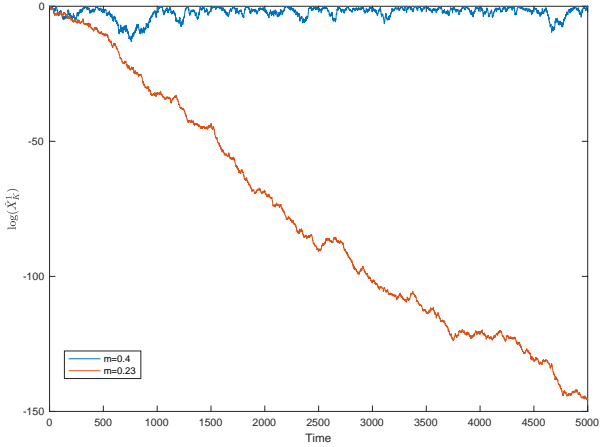}
\caption{Example of the super-martingale phenomenon. Pre and post change densities are unit variance Gaussians with means of 0 and $m$, respectively; $\rho=0.05$. The super-martingale phenomenon emerges when the measurements are insufficiently informative ($m=0.23$) 
%red curve with post-change mean $m=0.23$) 
compared to more reasonable behaviour when the measurements are informative (%blue curve with post-change mean 
$m=0.4$).}
\label{fig:gauss}
\end{center}
\end{figure}

\paragraph*{Simulation:}
Consider a geometric prior $\rho=0.05$ and note from Proposition \ref{lem:setgaussuan} that the phenomenon emerges below $m_c=0.32$. %$\sqrt{2\log(1/(1-\rho))}=0.32$.
%We will now present a illustrative simulation example of the %to illustrate the 
%discord monotonic posterior behaviour that emerges when $m \in E(\rho,\delta)$. % (at $m_c=\sqrt{2\log(1/(1-\rho))}$).
% simulate a sequence not containing a change event.
% % %\subsection{True model known case}
% % What happens on 10000 step non change sequence for different values of $b$ (change occurs at 10000). Assume filter correctly knows $b$ case. 
% Note that $\rho=0.05$ and represent significant mismatch in the filter's geometric prior (this high level of mismatch tend to emphasis the issues).  Here 
Figure \ref{fig:gauss} illustrates two simulated examples of the posterior's behaviour on a sequence prior to the change time ($m=0.40$ and $m=0.23$ representing examples of informative and non-informative measurements). The significantly different behaviour seen is an illustration of the super-martingale phenomenon discussed in this paper.
%in failure and not failure case (significant differences).
% %
% We also conducted a small 10 case Monte Carlo study with each of 
% $m \in \{0.23,0.31,0.4\}$. The phenomenon emerged for in all 10 cases of $m=0.23$, in 9 of 10 cases of $m=0.31$, and 0 of 10 cases of $m=0.40$. 
%Hence, whilst the $\hat{X}_{k }^1 < h_s$ interval  trap  is not always entered, 
%%(particular when near $m_c$ where the interval trap  will be smaller),
%the discord phenomenon is far from rare in this Gaussian example.

\begin{figure}
\begin{center}
\includegraphics[trim={0.5cm 0.5cm 0.5cm 0.5cm},clip,scale=0.45]{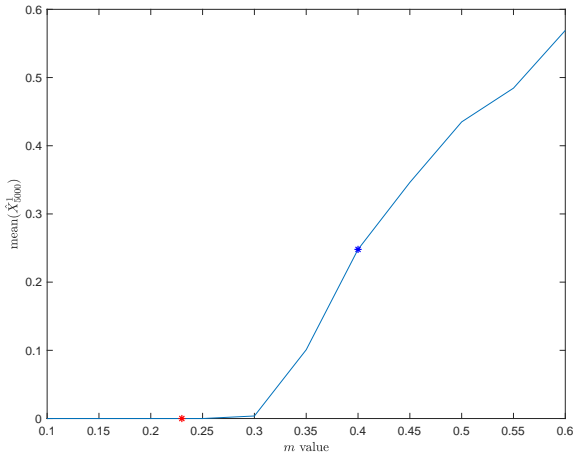}
\caption{Illustration of behaviour transition at the critical value. Mean value of $\hat{X}^1_{5000}$ when no change event (1000 trials, $\rho=0.05$, $m_c=0.32$). The stars mark the $m$ value cases studied in Figure \ref{fig:gauss}. }
\label{fig:monte}
\end{center}
\end{figure}

To illustrate %of the paper's observation about
%the domain and problematical nature of the %Shiryaev's test statistic's 
the  transition in the behaviour of Shiryaev's test statistic, % at the critical value $m_c$,
%super martingale phenomenon 
for each value of $m=0.1, 0.15, \ldots, 0.6$ we conducted a Monte-Carlo study of 1000 trials of 5000 long random variable sequences with no change. %in the absence of a change event
  In Figure \ref{fig:monte} the %Monte-Carlo mean value of no-change posterior %after 5000 steps 
mean value of $\hat{X}^1_{5000}$ ($\rho=0.05$)  illustrates that
% when above $m_c=0.32$ Shiryaev's test statistic provides a %(probabilistically interpretable) 
% (probabilistic) indication that a change has not occurred, whilst when below $m_c$ the test statistic becomes %(incorrectly) increasingly 
% incorrectly convinced that a change has occurred. 
below the critical value $m_c=0.32$ the test statistic exhibits the super-martingale phenomenon and
becomes %(incorrectly) increasingly 
 incorrectly convinced that a change has occurred, when it has not. 

\section{Discussion}
The super-martingale phenomenon emerges in Bayesian QCD as a consequence of the non-ergodic nature of the underlying signal model.
%and $\hat{X}_k^1 < h_\delta$ occurs probabilistically under $E_\pi$ due to the geometric prior.
%of Shiryaev's rule}.
% , and disappears in 
% %considering the modified Bayesian 
% the quickest intermittent signal detection problem \cite{James2018} which generalises Shiryaev's problem.
% \item   
That the class of post-change densities exhibiting the phenomenon 
%in our example problem 
can by parameterized by an interval set suggests this 
%phenomenon 
is a systemic issue of the problem rather than the result of a pathological noise realisation.
% Our simulation study illustrated that  the $\hat{X}_{k }^1 < h_s$ interval  trap  may not always be entered, 
% %(particular when near $m_c$ where the interval trap  will be smaller),
% but the phenomenon is far from rare.
% Potential remedies in applications with weak measurements involve subtle model adjustments, such as the  quickest intermittent signal detection problem \cite{James2018} which generalizes Shiryaev's problem, or using non-Bayesian QCD such as the Lorden criterion \cite{Lorden}. 
Potential remedies in applications with weak measurements include
%involve subtle model adjustments, such as the  
using quickest intermittent signal detection \cite{James2018} 
%which generalizes Shiryaev's problem, 
or using non-Bayesian QCD such as the Lorden criterion \cite{Lorden}. 
 % (e.g. having Markov chains with absorbing states).
Finally, we would expect similar phenomenon to arise in more complex Bayesian QCD or filter problems involving non-ergodic models with weak observations.

% \paragraph*{Remarks}
% \begin{enumerate}%[i)]
% % \item The significant different posterior (illustrate in the simulation example) appears suddenly with decreasing $m$ (below some $m_c \ge \sqrt{(2\log(1/(1-\rho))-\delta)}$) and hence the term ``discord" in this problem.
% \item The discord emerges here as a consequence of the non-ergodic nature of the underlying signal model.
% % , and disappears in 
% % %considering the modified Bayesian 
% % the quickest intermittent signal detection problem \cite{James2018} which generalises Shiryaev's problem.
% % \item   
% Potential remedies in applications with weak measurements involve subtle model adjustments, such as the  quickest intermittent signal detection problem \cite{James2018} which generalizes Shiryaev's problem, or using non-Bayesian QCD such as the Lorden criterion \cite{Lorden}.
% \item We would expect the similar issue to arise in filter problems involving non-ergodic HMM models with weak observations (e.g. having Markov chains with absorbing states).
% \end{enumerate}

% \section{Conclusion}
% \hl{ to be done}

%%%%%%%%%%%%%%%%%%%%%%%%%%%%%%%%%%%%%%%%%%%%%%%%%%%%%%%%%%%%%%%%%%%%%%%%%%%%%%%%%%%%%%%%%
 \paragraph*{Acknowledgment.} The authors express many thanks to an anonymous reviewer who helped correct the proof of Lemma 2.

%%%%%%%%%%%%%%%%%%%%%%%%%%%%5
%\bibliographystyle{IEEEtran}
\bibliographystyle{plain}    
\bibliography{IEEEabrv,ref}

\end{document}